\documentclass{JHEP}
\usepackage{epsfig,psfig,axodraw}

\newcommand{\be}{\begin{equation}}
\newcommand{\ee}{\end{equation}}
\newcommand{\br}{\begin{eqnarray}}
\newcommand{\er}{\end{eqnarray}}
\newcommand{\ba}{\begin{array}}
\newcommand{\ea}{\end{array}}
\newcommand{\bi}{\begin{itemize}}
\newcommand{\ei}{\end{itemize}}
\newcommand{\bn}{\begin{enumerate}}
\newcommand{\en}{\end{enumerate}}
\newcommand{\bc}{\begin{center}}
\newcommand{\ec}{\end{center}}

\newcommand{\ar}{\rightarrow}

\newcommand{\Dir}{\kern -6.4pt\Big{/}}
\newcommand{\Dirin}{\kern -10.4pt\Big{/}\kern 4.4pt}
\newcommand{\DDir}{\kern -10.6pt\Big{/}}
\newcommand{\DGir}{\kern -6.0pt\Big{/}}
\def\Ecm{\ifmmode{E_{\mathrm{cm}}}\else{$E_{\mathrm{cm}}$}\fi}
\def\gluino{\ifmmode{\mathaccent"7E g}\else{$\mathaccent"7E g$}\fi}
\def\photino{\ifmmode{\mathaccent"7E \gamma}\else{$\mathaccent"7E \gamma$}\fi}
\def\mgluino{\ifmmode{m_{\mathaccent"7E g}}
             \else{$m_{\mathaccent"7E g}$}\fi}
\def\taugluino{\ifmmode{\tau_{\mathaccent"7E g}}
             \else{$\tau_{\mathaccent"7E g}$}\fi}
\def\mphotino{\ifmmode{m_{\mathaccent"7E \gamma}}
             \else{$m_{\mathaccent"7E \gamma}$}\fi}
\def\ML{\ifmmode{{\mathaccent"7E M}_L}
             \else{${\mathaccent"7E M}_L$}\fi}
\def\MR{\ifmmode{{\mathaccent"7E M}_R}
             \else{${\mathaccent"7E M}_R$}\fi}

\def\jp #1 #2 #3 {{J.~Phys.} {#1} (#2) #3}
\def\pl #1 #2 #3 {{Phys.~Lett.} {#1} (#2) #3}
\def\np #1 #2 #3 {{Nucl.~Phys.} {#1} (#2) #3}
\def\zp #1 #2 #3 {{Z.~Phys.} {#1} (#2) #3}
\def\pr #1 #2 #3 {{Phys.~Rev.} {#1} (#2) #3}
\def\prl #1 #2 #3 {{Phys.~Rev.~Lett.} {#1} (#2) #3}
\def\mpl #1 #2 #3 {{Mod.~Phys.~Lett.} {#1} (#2) #3}
\def\rmp #1 #2 #3 {{Rev. Mod. Phys.} {#1} (#2) #3}
\def\sjnp #1 #2 #3 {{Sov. J. Nucl. Phys.} {#1} (#2) #3}
\def\cpc #1 #2 #3 {{Comp. Phys. Comm.} {#1} (#2) #3}
\def\xx #1 #2 #3 {{#1}, (#2) #3}
\def\Ord{\lower .7ex\hbox{$\;\stackrel{\textstyle <}{\sim}\;$}}
\def\OOrd{\lower .7ex\hbox{$\;\stackrel{\textstyle >}{\sim}\;$}}

 \title{
Improving the discovery potential of \\[0.25 cm]
charged Higgs bosons at Tevatron Run 2}

 \author{ 
Monoranjan Guchait and Stefano Moretti\\
CERN -- Theory Division, CH-1211 Geneva 23, Switzerland\\
Emails: monoranjan.guchait@cern.ch, stefano.moretti@cern.ch}

 \abstract{
By exploiting the full process $p\bar p\to t\bar b H^-$, 
in place of the standard Monte Carlo procedure of 
factorising production and decay, $p\bar p\to t\bar t$ 
followed by $\bar t\to\bar b H^-$, 
we show how to improve the discovery reach of the Tevatron (Run 2)
in charged Higgs boson searches, in the large
$\tan\beta$ region. This is achieved in conjunction with dedicated
cuts on a `transverse mass' distribution sensitive to the
Higgs boson mass and to `polarisation' effects in the 
$H^-\to\tau^-\bar\nu_\tau$ decay channel.}


 \preprint{
{CERN-TH/2001-243}\\
December 2001}

\clearpage

\begin{document}

\section{Charged Higgs bosons at the Tevatron}
\label{sec_intro}

The importance of charged Higgs boson searches at future colliders
has in the recent years been emphasised more and more  
\cite{conveners}--\cite{reviewLHC}: the detection
of a `scalar charged' particle would in fact definitely signal the existence
of New Physics beyond the Standard Model (SM). Such states are
naturally accommodated in non-minimal Higgs scenarios, such as
Two-Higgs Doublet Models (2HDMs). A Supersymmetric version of the
latter is the Minimal Supersymmetric Standard Model (MSSM) -- in fact,
a Type II 2HDM with specific relations among neutral
and charged Higgs boson masses and couplings, as dictated by
Supersymmetry (SUSY) \cite{guide}. 

The
Tevatron collider at Fermilab has just begun its second stage
of operation, so-called
Run 2, with a higher centre-of-mass (CM) energy ($\sqrt s=2$
TeV) and a prospect of collecting something like 15 fb$^{-1}$ of
luminosity (per experiment) by the end of its lifetime. This machine
will be the first one to probe charged Higgs boson masses in the 
mass range $M_{H^\pm}\sim m_t$ \cite{reviewTEV}. At present, a lower bound
on the charged Higgs boson mass exists from LEP \cite{LepTre},
$M_{H^\pm}\OOrd M_{W^\pm}$,
independently of the charged Higgs branching ratios (BRs).
This limit is valid within a general 2HDM whereas, in 
the low $\tan\beta$ region (say, below 3), an indirect lower 
limit on $M_{H^\pm}$ can be derived  in the MSSM 
from the one on $M_{h^0}$ (the mass
of the lightest Higgs state of the model):
$M_{H^\pm}^2\approx M_{W^\pm}^2+M_{h^0}^2\OOrd (130~\mathrm{GeV})^2$.

The main production mode of $H^\pm$ scalars at the Tevatron, for
$M_{H^\pm}<m_t$, is the decay of top
(anti)quarks, the latter being produced via QCD in the annihilation
of gluon-gluon and quark-antiquark pairs. 
Simulation studies aiming to assess the discovery reach of the Tevatron
in the quest for charged Higgs bosons have relied so far
on Monte Carlo (MC) programs, such as PYTHIA \cite{pythia}, HERWIG 
\cite{herwig} and ISAJET \cite{isajet}. Here, the above process is
accounted for through the usual procedure of factorising
the production process, $gg,q\bar q\to t\bar t$, times the
decay one, $\bar t\to \bar b H^-$, in the so-called on-shell
top approximation, 

It is the purpose of this letter to show how this description
fails to correctly describe the production and decay 
phenomenology of charged Higgs bosons when their mass approaches the top one, 
hence undermining the ability of experimental analyses at Tevatron
in pinning down
the real nature of these particle (if not detecting them altogether). 
We will do so by comparing the results obtained in the above approximation 
with those produced through the full processes
$g_1g_2,q_1\bar q_2\to t_3\bar b_4 H^-_5$  \cite{tbH},
proceeding via the following graphs:
\begin{eqnarray}
 \nonumber
\hskip-0.5cm{{G}_{1}=}
\begin{picture}(150,50)
\SetScale{1.0}
\SetWidth{1.2}
\SetOffset(0,-60)
\Gluon(45,75)(30,90){3}{3}
\Gluon(30,40)(45,55){3}{3}
\Text(25,95)[]{\small 1}
\Text(25,35)[]{\small 2}
\ArrowLine(60,40)(45,55)
\ArrowLine(45,55)(45,75)
\ArrowLine(45,75)(60,90)
\DashLine(55,85)(70,85){2}
\Text(77.5,85)[]{\small 5}
\Text(65,35)[]{\small 4}
\Text(65,95)[]{\small 3}
\end{picture} 
{G}_{2}=
\begin{picture}(150,50)
\SetScale{1.0}
\SetWidth{1.2}
\SetOffset(0,-60)
\Gluon(45,75)(30,90){3}{3}
\Gluon(30,40)(45,55){3}{3}
\Text(25,95)[]{\small 1}
\Text(25,35)[]{\small 2}
\ArrowLine(60,40)(45,55)
\ArrowLine(45,55)(45,75)
\ArrowLine(45,75)(60,90)
\DashLine(45,65)(60,65){2}
\Text(67.5,65)[]{\small 5}
\Text(65,35)[]{\small 4}
\Text(65,95)[]{\small 3}
\end{picture} 
{G}_{3}=
\begin{picture}(150,50)
\SetScale{1.0}
\SetWidth{1.2}
\SetOffset(0,-60)
\Gluon(45,75)(30,90){3}{3}
\Gluon(30,40)(45,55){3}{3}
\Text(25,95)[]{\small 1}
\Text(25,35)[]{\small 2}
\ArrowLine(60,40)(45,55)
\ArrowLine(45,55)(45,75)
\ArrowLine(45,75)(60,90)
\DashLine(55,45)(70,45){2}
\Text(77.5,45)[]{\small 5}
\Text(65,35)[]{\small 4}
\Text(65,95)[]{\small 3}
\end{picture} 
\\ \nonumber
\\ \nonumber
\\ \nonumber
\hskip-0.5cm{{G}_{4}=}
\begin{picture}(150,50)
\SetScale{1.0}
\SetWidth{1.2}
\SetOffset(0,-60)
\Gluon(45,75)(30,90){3}{3}
\Gluon(30,40)(45,55){3}{3}
\Text(25,95)[]{\small 2}
\Text(25,35)[]{\small 1}
\ArrowLine(60,40)(45,55)
\ArrowLine(45,55)(45,75)
\ArrowLine(45,75)(60,90)
\DashLine(55,85)(70,85){2}
\Text(77.5,85)[]{\small 5}
\Text(65,35)[]{\small 4}
\Text(65,95)[]{\small 3}
\end{picture} 
{G}_{5}=
\begin{picture}(150,50)
\SetScale{1.0}
\SetWidth{1.2}
\SetOffset(0,-60)
\Gluon(45,75)(30,90){3}{3}
\Gluon(30,40)(45,55){3}{3}
\Text(25,95)[]{\small 2}
\Text(25,35)[]{\small 1}
\ArrowLine(60,40)(45,55)
\ArrowLine(45,55)(45,75)
\ArrowLine(45,75)(60,90)
\DashLine(45,65)(60,65){2}
\Text(67.5,65)[]{\small 5}
\Text(65,35)[]{\small 4}
\Text(65,95)[]{\small 3}
\end{picture} 
{G}_{6}=
\begin{picture}(150,50)
\SetScale{1.0}
\SetWidth{1.2}
\SetOffset(0,-60)
\Gluon(45,75)(30,90){3}{3}
\Gluon(30,40)(45,55){3}{3}
\Text(25,95)[]{\small 2}
\Text(25,35)[]{\small 1}
\ArrowLine(60,40)(45,55)
\ArrowLine(45,55)(45,75)
\ArrowLine(45,75)(60,90)
\DashLine(55,45)(70,45){2}
\Text(77.5,45)[]{\small 5}
\Text(65,35)[]{\small 4}
\Text(65,95)[]{\small 3}
\end{picture} 
\end{eqnarray}
\vskip-0.2cm
\begin{eqnarray}\label{QQgg}
\nonumber
\hskip-3.25cm{{G}_{7}=}
\begin{picture}(150,50)
\SetScale{1.0}
\SetWidth{1.2}
\SetOffset(2,-62.5)
\Gluon(15,50)(30,65){3}{3}
\Gluon(30,65)(15,80){3}{3}
\Text(10,45)[]{\small 2}
\Text(10,85)[]{\small 1}
\Gluon(30,65)(60,65){3}{3}
\ArrowLine(75,50)(60,65)
\ArrowLine(60,65)(75,80)
\DashLine(70,75)(85,75){2}
\Text(92.5,75)[]{\small 5}
\Text(80,45)[]{\small 4}
\Text(80,85)[]{\small 3}
\end{picture} 
~{G}_{8}=
\begin{picture}(150,50)
\SetScale{1.0}
\SetWidth{1.2}
\SetOffset(2,-62.5)
\Gluon(15,50)(30,65){3}{3}
\Gluon(30,65)(15,80){3}{3}
\Text(10,45)[]{\small 2}
\Text(10,85)[]{\small 1}
\Gluon(30,65)(60,65){3}{3}
\ArrowLine(75,50)(60,65)
\ArrowLine(60,65)(75,80)
\DashLine(70,55)(85,55){2}
\Text(92.5,55)[]{\small 5}
\Text(80,45)[]{\small 4}
\Text(80,85)[]{\small 3}
\end{picture}
\\ \nonumber
\end{eqnarray}
\begin{eqnarray}\label{procs}
\hskip-3.25cm{{Q}_{1}=}
\begin{picture}(150,50)
\SetScale{1.0}
\SetWidth{1.2}
\SetOffset(2,-62.5)
\ArrowLine(30,65)(15,50)
\ArrowLine(15,80)(30,65)
\Text(10,45)[]{\small 2}
\Text(10,85)[]{\small 1}
\Gluon(30,65)(60,65){3}{3}
\ArrowLine(75,50)(60,65)
\ArrowLine(60,65)(75,80)
\DashLine(70,75)(85,75){2}
\Text(92.5,75)[]{\small 5}
\Text(80,45)[]{\small 4}
\Text(80,85)[]{\small 3}
\end{picture} 
~{Q}_{2}=
\begin{picture}(150,50)
\SetScale{1.0}
\SetWidth{1.2}
\SetOffset(2,-62.5)
\ArrowLine(30,65)(15,50)
\ArrowLine(15,80)(30,65)
\Text(10,45)[]{\small 2}
\Text(10,85)[]{\small 1}
\Gluon(30,65)(60,65){3}{3}
\ArrowLine(75,50)(60,65)
\ArrowLine(60,65)(75,80)
\DashLine(70,55)(85,55){2}
\Text(92.5,55)[]{\small 5}
\Text(80,45)[]{\small 4}
\Text(80,85)[]{\small 3}
\end{picture}
\\ \nonumber
\end{eqnarray}
\vskip0.5cm\noindent
In fact, we will argue the latter being the correct way to describe 
charged Higgs boson production and decay in the `threshold
region': $M_{H^\pm}\sim m_t$. Specifically, we will be exploring
Higgs mass values beyond the customary 160 GeV limit considered
in Run 2 studies \cite{reviewTEV}, while remaining below 190 GeV, where
the production cross section is below detection level.

Finally, we will proceed to a signal-to-background analysis, the latter 
incorporating dedicated selection procedures already advocated in literature, 
in order to illustrate how the $H^\pm$ discovery potential of the Tevatron 
can be improved, in the context of so-called `direct' (or `appearance') 
searches \cite{CDFdir}. In such a case, specific decay modes of charged
Higgs bosons are searched for and kinematical selections are optimised
to extract one or another decay signature. In contrast, in
`indirect' (or `disappearance') searches \cite{CDFD0ind}, one
employs selection criteria optimised to detect the SM decay
of a top (anti)quark, $t\to bW^+\to b X$, and any loss of such events
can be ascribed to the presence of $t\to bH^+\to b X$ decays. 
As remarked in \cite{reviewTEV}, the latter method is expected
to yield stronger (null) results for integrated luminosities below
2 fb$^{-1}$ or so, whereas with increasing statistics (and, possibly,
enhanced detector performances) the former is expected to dominate.
Conversely, if a charged Higgs boson exists with mass around $m_t$,
its presence could be detected through a disappearance search, hence prompting
a direct search to confirm discovery.

In this note, among the possible decay signatures of $H^\pm$ states,
we will concentrate on the $H^-\to \tau^-\bar\nu_\tau$ channel, which is the
dominant one for our considered range of $M_{H^\pm}$. The 
$H^-\to b\bar t$ signature originating from $t\bar b H^-$ final states
has already been considered (for relatively higher $M_{H^\pm}$ values) in 
Ref.~\cite{Jaume}. For a review of typical decay rates of charged
Higgs bosons, see Ref.~\cite{BRs}.

\section{Charged Higgs boson production in the threshold region}
\label{sec_threshold}

The subprocesses in (\ref{procs}) account for both top-antitop
production and decay (graphs $G_3,G_6$ and $G_8$ for gluon-gluon
and $Q_2$ for quark-antiquark) as well as for Higgs-strahlung
(all other graphs) and the relative interferences. In fact, in order
to emulate the current implementation in MC programs,
one can extract the top-antitop graphs in a gauge invariant fashion,
by setting $G_1=G_2=G_4=G_5=G_7=0$ plus $Q_1=0$, and rewriting,
in the fixed width scheme,
the top propagator as (here, $p=p_4+p_5$)
\begin{equation}\label{prop}
\frac{ p\Dir  + m_t}{p^2-m^2_t+im_t\Gamma_t}
\left( \frac{\Gamma_t}{\Gamma_{\mathrm{tot}}}\right)^{\frac{1}{2}}.
\end{equation}
When $\Gamma_t = \Gamma_{\mathrm tot}$, the total top width,
the standard expression is recovered. The on-shell or Narrow
Width Approximation (NWA) can be obtained by
taking numerically $\Gamma_t \ar 0$, as in this limit
eq.~(\ref{prop}) becomes a representation of the Dirac
delta function $\delta(p^2-m_t^2)$ (apart from a factor
$\pi$). In practice, the cross section coincides with
the one computed as production times BR already for 
$\Gamma_t \sim  10^{-3}$.
For the width of the top-quark 
we have used the tree-level expression, which depends
upon both $M_{H^\pm}$ and $\tan\beta$
(the ratio between the vacuum expectation values of
the two doublet Higgs fields). Similarly, we have proceeded with the
couplings entering the $t\bar b H^-$ vertex.

In Fig.~\ref{fig_threshold}, we compare the total cross section
obtained by computing processes (\ref{procs}) (with 
$\Gamma_t = \Gamma_{\rm{tot}}$) to the $t\bar t$-mediated
one in NWA, i.e.,
$gg,q\bar q\to t\bar t\to t\bar b H^-$ with $\Gamma_t \ar 0$.
These results are both gauge invariant. For the sake of illustration,
we also have included here similar rates obtained by computing the
top-antitop diagrams only, with $\Gamma_t=\Gamma_{\mathrm{tot}}$
(these are subject to a gauge dependence of order
${\cal O}(\Gamma_t/m_t)$)\footnote{They are meant to
illustrate what portion of the difference between the full results and
those in NWA is due to finite width effects of the top-quark, the
remainder of it coming from the contribution of the other diagrams
and the relative interferences.}. Cross sections are shown for two 
representative $\tan\beta$ values (3 and 30) over the mass range
$160~{\mathrm{GeV}}~< M_{H^\pm} < 190$ GeV. However, since the effect
that we are investigating is merely kinematical, the same quantitative
features would appear for other choices of the former. It is evident 
how, with $M_{H^\pm}$ approaching $m_t$, the Higgs boson rates are
grossly mis-estimated by the NWA. Hence, it is mandatory to exploit
in future MC simulations of $H^\pm$ production and decay 
around the threshold region an implementation based on the
$gg,q\bar q\to t\bar b H^-$ matrix elements. Besides, for
$M_{H^\pm}>m_t-m_b$, only the latter can produce a non-zero result.
Our default values for $m_t$ and $m_b$ are 175 and 4.25 GeV,
respectively,
both in the couplings and kinematics\footnote{When we will
discuss decay rates of the charged Higgs boson, the two masses above will be 
the input values of HDECAY \cite{HDECAY} -- the package that we used for 
our numerical estimates.}.

Also differential distributions can strongly be affected by an
approximated modelling of the production process in the
threshold region. In Fig.~\ref{fig_pt}, we present the spectra
in transverse momentum of the top and bottom quarks, 
for the full $2\to3$ process and the NWA. Whereas differences in $p_T$
are negligible in the case of the Higgs boson (so that this case
is not plotted here), they
are sizable for the top and dramatic for the bottom quark. Whereas
one should expect the impact of the differences seen in the top quark
distribution to eventually be marginal, owning to the fact that this
particle is actually unstable and that its three-body decay products are 
subject to 
the (cumulative) effect of usual detector resolution uncertainties, this
is no longer true for the bottom quark, which fragments directly
into hadrons. Besides,
the availability of the newly implanted silicon vertex detector  
may render the tagging of $b$-quarks a crucial ingredient in detection 
strategies of charged Higgs bosons at Run 2.
Results in Fig.~\ref{fig_pt} are shown for $M_{H^\pm}=170$ GeV
and $\tan\beta=3$. Whereas the described effects are insensitive to
the actual value of the latter, a difference choice of the former
can modify the relative shape of the two curves
(full and NWA) significantly, but the distinctive features seen here 
remain qualitatively the same for any choice of $M_{H^\pm}$ in
the considered mass interval.

Before moving on to the decay analysis, one final
remark is in order concerning the production stage. 
In Fig.~\ref{fig_threshold}, we also have 
presented the cross sections of the 
so-called `Drell-Yan mode', $q\bar q\to H^+H^-$ \cite{qqHH}\footnote{In fact, 
this nomenclature is somewhat misleading, as $b\bar b\to
H^+H^-$ contributions proceeding via double Higgs-strahlung or
neutral Higgs boson mediation (e.g., $h^0,H^0$ and $A^0$ in the MSSM),
as opposed to gauge boson exchange, i.e., $q\bar q\to \gamma^*,Z^*\to
H^+H^-$, are not entirely negligible, particularly at the LHC, 
where they are in fact dominant at large $\tan\beta$ -- similarly for the
loop-induced contributions, $gg\to H^+H^-$ \cite{ggHH}.
(Notice that we include all such subprocesses in our calculation.)},
followed by $H^+\to t\bar b$. This is the only
process competing with $gg,q\bar q\to t\bar b H^-$ 
at Tevatron energies, at least in the low to intermediate $\tan\beta$
region, say, $1\Ord\tan\beta\Ord10$. (At the Large Hadron
Collider (LHC), many more production modes exist \cite{LHC}.)
In principle then, one should also investigate the $q\bar q\to H^+H^-\to
t\bar b H^-$ channel, alongside $gg,q\bar q\to t\bar b H^-$
in (\ref{procs}). 
In practice, though, in direct investigations
of the $H^-\to \tau^-\bar\nu_\tau$ decay channel, one is implicitly
concerned with large $\tan\beta$ values only. The reason is twofold.
On the one hand, an appearance search in the threshold region with a 
very low $\tan\beta$ (say, below 1.5) would have to be based on 
$H^-\to s\bar c$ decays, which are very challenging because of an 
overwhelming QCD noise; whereas for $\tan\beta=2$--3 one finds that 
$H^\pm\to W^{\pm(*)} h^0$ decays can be relevant (see \cite{Wh} for some 
LHC studies), and these are strongly model dependent  (e.g., in the MSSM
they are no longer viable, given the recent limits on $M_{h^0}$ from LEP
in this scenario). To date, CDF \cite{CDFdir} has only published results
for $H^-\to\tau^-\bar\nu_\tau$, which are valid for $\tan\beta\OOrd4$.
On the other hand, for $\tan\beta$ close to $\sqrt {m_t/m_b}$,
the strength of the $t\bar b H^-$ coupling -- entering the diagrams
in (\ref{procs}) -- reaches a minimum, in the end rendering the production
cross section unobservable. This basically occurs over the 
interval $4\Ord\tan\beta\Ord10$. Besides, for $M_{H^\pm}\sim m_t$ and 
$\tan\beta\OOrd10$, the $H^-\to \tau^-\bar\nu_\tau$ decay mode is truly 
dominant, as one can appreciate by combining the production cross sections of 
Fig.~\ref{fig_threshold} with typical $H^\pm$ decay rates \cite{BRs}.
This is made clear in the left-hand side of Fig.~\ref{fig_sigmaBRs},
for a general Type II 2HDM (note the relevance of $H^-\to b\bar t^{(*)}$
off-shell decays at low $\tan\beta$ even well below $M_{H^\pm}= m_t+m_b$). 
The statement remains true also in
the case of charged Higgs boson decays into Supersymmetric particles 
\cite{hdksusy},
as could well happen in the MSSM (according to current experimental limits).
For example, in the right-hand side of Fig.~\ref{fig_sigmaBRs},
we display the corresponding $\sigma\times {\mathrm{BR}}$ rates in the MSSM
with $M_2=130$ GeV and $\mu=300$ GeV, yielding:
$M_{{\tilde\chi}^{\pm}_{1,2}}\approx 330, 106$ GeV and
$M_{{\tilde\chi}^{\pm}_{1,2,3,4}}\approx 58, 109, 304, 331$ GeV
(for $\tan\beta=3$);
$M_{{\tilde\chi}^{\pm}_{1,2}}\approx 118, 326$ GeV and
$M_{{\tilde\chi}^{\pm}_{1,2,3,4}}\approx 63, 119, 309, 322$ GeV
(for $\tan\beta=30$). Here,
$M_{{\tilde\chi}^{\pm}_i}$ and $M_{{\tilde\chi}^{0}_j}$ 
are the two ($i=1,2$) chargino 
and four ($j=1,2,3,4$) neutralino masses, respectively. In fact,
the dominant decays into Supersymmetric particles of $H^\pm$
bosons are those into chargino-neutralino pairs, since typically one
has $M_{H^\pm}>M_{{\tilde\chi}^{\pm}_i} + M_{{\tilde\chi}^{0}_j}$
for some $ij$ combination (see Ref.~\cite{bisset},
where more plots along the same lines can also be found).
We defer the detailed investigation of the low to intermediate $\tan\beta$
interval to  \cite{preparation}.

\section{Signal selection in the $H^-\to \tau^-\bar\nu_\tau$ channel}
\label{sec_results}

The signature of interest here is $p \bar p \rightarrow t \bar b H^-$, 
followed by $H^- \rightarrow \tau^- \bar\nu_{\tau}$, with the top
quark decaying hadronically, $t \rightarrow b q \bar q'$. The same type 
of event topologies may appear 
in the SM process $p \bar p \rightarrow t \bar b W^-
\to t\bar b\tau^- \bar\nu_\tau$,
which is in fact the dominant irreducible background. This should be clear,
if one notices that one of the subprocesses entering the background is
$p \bar p \rightarrow t \bar t \rightarrow 
t \bar b W^-$, i.e., top-antitop production and decay in the SM,
for which one has $\sigma(p \bar p \rightarrow
t \bar t) \sim$ 7--8 pb at Tevatron for $\sqrt s=$2 TeV
(in our calculation, the full set of tree-level diagrams leading to
$t\bar b W^-$ final states has been computed).
 
The $\tau$'s can be tagged as narrow jets in their `one-prong' hadronic decay 
modes, which represent 90\% of the hadronic decay rate and about 
50\% of the total one. The main components of
such decays are: $\tau^\pm \rightarrow 
\pi^\pm \nu_\tau(12.5\%), \rho^\pm \nu_\tau(24\%)$ and $a_1^\pm 
\nu_\tau(7.5\%)$, with in turn $\rho^\pm\to\pi^\pm\pi^0$ and
$a_1^\pm\to\pi^\pm\pi^0\pi^0$. This distinguishing feature is in contrast
to the typical appearance of quark- and gluon-jets, which yield 
`multi-prong' hadronic topologies
in the detectors. This characteristic difference can profitably be 
exploited to efficiently isolate the hadronic $\tau$-signals from QCD
backgrounds of the form $W^\pm + {\mathrm {jets}}$ and 
$Z^0 + {\mathrm {jets}}$, which we have then ignored here.

We have studied our signal and background processes using a very simple
parton-level MC analysis, i.e., without taking into account
fragmentation effects of partons.  In our numerical calculation 
we have set the renormalisation and factorisation scales equal to 
the partonic CM energy,
$Q^2={\hat{s}}$, and the CTEQ4M \cite{cteq4} Parton Distribution Functions
 were used throughout.
For the selection of events, we have adopted the following set of cuts in
transverse momentum, $p_T$, and pseudorapidity, $\eta$, 
as well as transverse missing momentum, ${p\!\!\!/}_T$. 

\begin{enumerate}
\item
Tau-jets are selected if they satisfy the following criteria:
$p_T^{\tau} >$15 GeV and $|\eta^\tau|<$ 2.5. 
\item
We require ${p\!\!\!/}_T>$ 20 GeV, since the 
presence of neutrinos from $H^-$ decays and invisible decay products 
of $\tau$'s (mainly $\pi^0$'s) implies that a significant fraction of 
transverse momentum goes undetected.
\item
Quark-jets are selected by imposing $p_T^j >$ 20 GeV and $|\eta^j| <$ 2.5.
We require at least one of these to be tagged as a $b$-jet. 
\item 
We demand that two un-tagged jets have an invariant mass
around $M_{W^\pm}$, e.g., $|M_{q\bar q'} - M_{W^\pm}| <$ 10 GeV
and that the $b$-jet in combination with other two un-tagged jets produces
an invariant mass close to $m_t$, e.g., $|M_{b q \bar q'} - m_t| <$
15 GeV. 
\end{enumerate}
After the implementation of these cuts, we have found that
the cross section for the signal is, e.g., 0.6(5.5) fb for $M_{H^\pm} =$170 
GeV and $\tan\beta=3(40)$; whereas for the background process one has
90 fb. Clearly, signal-to-background ratios
($S/B$'s) of this sort are insufficient to establish
the presence of $H^\pm$ states. Hence, further cuts have to be devised.

To this end, we have exploited another kinematic
variable: a transverse mass, $M_T$, constructed from the visible
$\tau$-jet and the missing energy, i.e.,
\begin{equation}\label{mT} 
M_T =\sqrt{2 p_T^{\tau}
{p\!\!\!/}_T (1 - \cos\Delta\phi)},
\end{equation}
as introduced in Ref.~\cite{mTr}. In the case of the signal, the 
$\tau$-jets are heavily boosted relatively to the case of the background, 
as the charged Higgs masses considered here are much heavier 
than $M_{W^\pm}$. This leads to 
a backward(forward) peak in the azimuthal angle distribution, $\Delta \phi$, 
identified by the directions of the $\tau$-jet and the 
missing momentum in the signal(background): see Fig.~\ref{fig_mT}. 
By imposing  $M_T >M_{W^\pm}\approx$ 80 GeV, the background is reduced by 
more than two orders 
of magnitude, while the signal cross section is suppressed to a 
much lesser extent. For example, for $M_{H^\pm}=$170 GeV, the latter
becomes 0.4(3.5) fb for $\tan\beta=$ 3(40) while the former comes down 
to a manageable 0.22 fb.  In Tab.~1 we summarise the signal and background 
cross sections for some representative values
of $M_{H^\pm}$ and $\tan\beta$, after all cuts described 
above. 

Before converting the numbers in Tab.~1 into event rates
and significances, one has to take into account the finite efficiency
of the detectors in particle identification. For example, 
$\tau$-identification efficiencies are estimated to be of order 
 50\% \cite{tauid}, similarly for the tagging of any $b$-jet \cite{reviewTEV}. 
Hence, one should more realistically expect both signal and background rates 
to be further reduced by a factor of 4 or so. In the end, however,
the chances of extracting the $H^\pm\to\tau^\pm\nu_\tau$ signal 
after 15 fb$^{-1}$ of luminosity are rather
good, at least at large $\tan\beta$, while being negligible at low to 
intermediate values of the latter (as already argumented). 
Notice that this remains true for charged 
Higgs masses above $m_t$ as well, say, up to 180 GeV or so, where a handful 
of signal events should survive in each experiment.  
 
This situation is rather encouraging, especially considering that
there may be some room to further improve the $S/B$'s if one recalls that
the distributions of one-prong hadronic decay tracks of 
$\tau$'s are strongly sensitive to the polarisation state of the lepton 
(see, e.g., Refs.~\cite{hagiwara,dptau} for a detailed discussion). 
Basically, the key feature relevant to our purposes is the correlation
between the latter and the energy sharing among the decay pions.
In fact, it is to be noted that the spin
state of $\tau$'s coming from $H^\pm$- and $W^\pm$-boson
decays are opposite: i.e., $H^- \rightarrow \tau^-_R \bar\nu_R$ and $
H^+ \rightarrow \tau^+_L \nu_L$ whereas
$W^- \rightarrow \tau^-_L \bar\nu_R$ and $W^+ \rightarrow \tau^+_R \nu_L$
(neglecting leptonic mass effects, as we did here).
Ultimately, this leads to a significantly harder momentum
distribution of charged pions from $\tau$-decays for the 
$H^\pm$-signal compared to the $W^\pm$-background, which can then be exploited
to increase  $S/B$. This is true
for the case of one-prong decays into both $\pi^\pm$'s and longitudinal
vector mesons, while the transverse component of the latter dilutes
the effect and must be somehow eliminated. This can be done inclusively, 
i.e., without having to identify the individual mesonic component 
of the one-prong hadronic topology. In doing so \cite{dpcode}, we will 
closely follow Ref.~\cite{dptau}.  

The mentioned transverse components of the signal as well as those 
of the background can adequately be suppressed
by requiring that 80\% of the $\tau$-jet (transverse) 
energy is carried away by 
the $\pi^\pm$'s, i.e.: 
\begin{equation}\label{frac}
R=\frac{p^{\pi^\pm}}{p_T^{\tau}}> 0.8.
\end{equation}
The enforcement of this constraint reduces by a factor of 5 
the background, while costing to the signal a 50\% suppression 
(for any relevant charged Higgs mass).
  
Incidentally, we should mention that acceptance efficiencies 
for the selection procedure that
we have chosen here are very similar in the case of the signal
for both the $2\to3$  simulation and the NWA. However,
this should not be surprising,
as we have imposed no requirement of a second $b$-tag. In fact,
in most cases, only one $b$-quark enters the detector
region -- the one produced in the (hadronic) decay of the
top quark in the $t\bar b H^-$ final state. In contrast,
things would be rather different if two $b$-tags were asked,
both at $p_T^b>20$ GeV, as it should be clear from the right-hand side
plot in Fig.~\ref{fig_pt}.

\section{Conclusions}

In summary, we have demonstrated that concrete prospects exist
at Tevatron Run 2 of extending the discovery reach of charged Higgs
bosons up to masses of order $m_t$, in the large $\tan\beta$ region,
in the context of direct searches.
This can be achieved by combining the following ingredients. 
\begin{itemize}
\item To emulate the production of charged Higgs boson events
by resorting to the full $gg,q\bar q\to t\bar b H^-$ process,
as opposed to the traditional procedure of generating the scalar
particles in on-shell top decays, from $gg,q\bar q\to t\bar t$
events. In fact, the former not only includes the dynamics of the
latter, but
also embeds charged Higgs production from Higgs-strahlung
and relative interferences.
\item To search for `one-prong' hadronic decays of $\tau$-leptons produced
in $H^-\to\tau^-\bar\nu_\tau$ events, in presence of a single $b$-tag,
 usual detector requirements and after $W^\pm$- and $t$-mass
reconstruction in the accompanying hadronic system, 
$t\to b W^+\to$~jets.
\item To enforce a cut in a typical transverse mass ($M_T$) distribution, which
is bound to assume values below the particle yielding the $\tau$-leptons
($H^\pm$ for the signal and $W^\pm$ for the background). 
Besides, since a cut as low as $M_T>M_{W^\pm}$ is sufficient to reduce the 
$W^\pm$ background to negligible levels, the same distribution can also be used
to eventually fit the unknown charged Higgs boson mass,
 when $M_{H^\pm}\sim m_t$.
\item Finally, to exploit well-known polarisation effects 
 in the case of $\tau^\pm\to \pi^\pm \nu_\tau$,
$\rho^\pm \nu_\tau$ and $a_1^\pm \nu_\tau$ decays.
\end{itemize}

In drawing our conclusions, we have relied on a parton-level analysis.
However, we expect that its main features should remain valid even in presence
of fragmentation/hadronisation effects. We do advocate a selection procedure
along the above lines to be investigated 
at a more phenomenological level (including realistic detector
simulations) by the Tevatron experiments. For example, the mentioned
$2\to3$ description of the $H^\pm$ production dynamics is 
available since version 6.3 in the HERWIG event generator
\cite{HW63} while polarised $\tau$-decays are now implemented 
in version  6.4 \cite{HW64}
(also including an interface to TAUOLA \cite{tauola}).

In order to motivate such analyses, we propose a benchmark scenario 
that may eventually emerge from these. In Fig.~\ref{fig_excl}, we 
present the exclusion regions, below the level curves,
in the $\tan\beta$--$M_{H^\pm}$ plane 
that can potentially be explored at 95\% confidence level (CL)
at the upgraded Tevatron, for two luminosity options, 5 and
15 fb$^{-1}$, by using the tools and the strategy outlined here. 
The significances $\sigma\equiv S/{\sqrt B}$ used for the contours
in Fig.~\ref{fig_excl} have been 
estimated at the parton level, after the sequence of cuts 
in 1.--4. and the one in transverse mass, $M_T >M_{W^\pm}$, hence before
the one in (\ref{frac}) and without including the mentioned efficiencies.
Incomplete as these estimates might be at this stage, it is clear the dramatic
improvement (both in $\tan\beta$ and $M_{H^\pm}$ reach) 
that could be achieved, if one compares our plot to
Fig.~102 of Ref.~\cite{reviewTEV}.

\section*{Acknowledgements}

MG is thankful to the High Energy Group at the
Abdus Salam International Centre for Theoretical Physics for hospitality
during the final phase of this work.

\vfill\clearpage

\FIGURE[!t]{
\epsfig{file= newthreshold3.ps,angle=90,height=5.cm}
\epsfig{file=newthreshold30.ps,angle=90,height=5.cm}\\
\caption{Cross section for $gg,q\bar q\to t\bar b H^-$ (solid),
$gg,q\bar q\to t\bar t\to t\bar b H^-$ (dashed, with finite top width) and
$gg,q\bar q\to t\bar t\to t\bar b H^-$ (dotted, in NWA), at $\sqrt s=2$ TeV,
as a function of $M_{H^\pm}$ for two representative values of $\tan\beta$
(hereafter, charge conjugated rates are always included).
For comparison, we  also have
plotted the cross section for $q\bar q\to H^+H^-\to
t\bar b H^-$ (dot-dashed). (Notice that the rates for the latter have been
multiplied by 10 for the case $\tan\beta=30$, for readability.)}
\label{fig_threshold}
}

\vfill\clearpage

\FIGURE[!b]{
\vspace{-1.0in}
\hspace{-1.0in}
\epsfxsize= 10.cm
\leavevmode
\epsffile{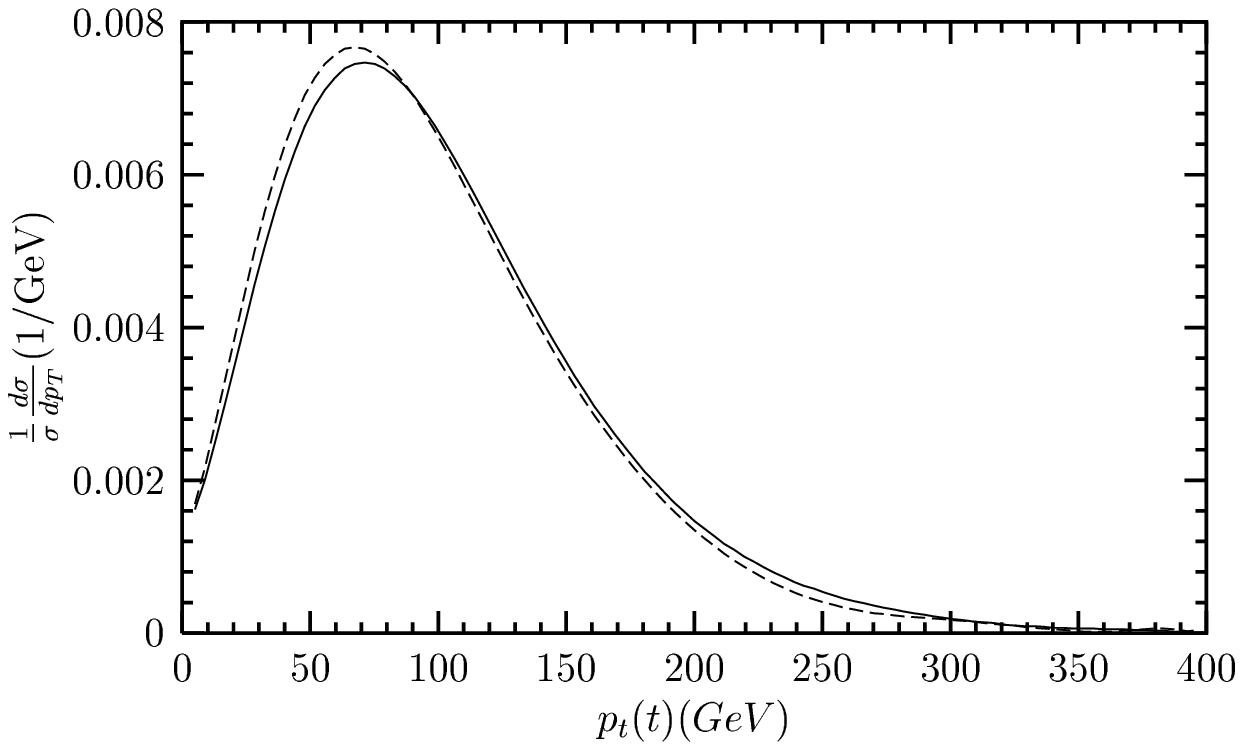}
\epsfxsize= 10.cm
\leavevmode
\hspace{-1.3in}
\epsffile{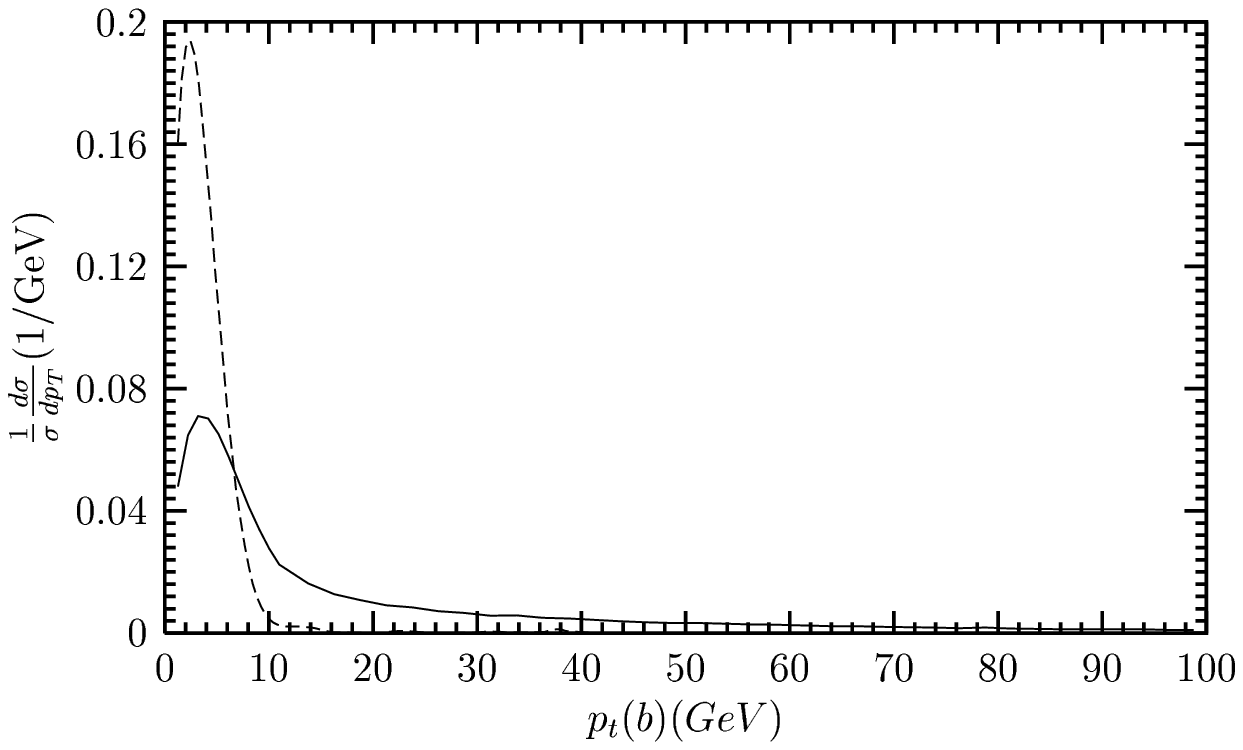}
\leavevmode
\vspace{-2.00in}
\caption{Transverse momentum distributions of the three-body
final state top- (left) and bottom-quark (right) 
in $gg,q\bar q\to t\bar b H^-$ (solid) and
$gg,q\bar q\to t\bar t\to t\bar b H^-$ (dashed) (the latter 
in NWA), at $\sqrt s=2$ TeV, for $M_{H^\pm}=170$ GeV.
 (Notice that the spectra are independent
of the choice of $\tan\beta$.)}
\label{fig_pt}
}

\FIGURE[!t]{
\vspace{-1.0in}
\hspace{-1.0in}
\epsfxsize= 10.cm
\leavevmode
\epsffile{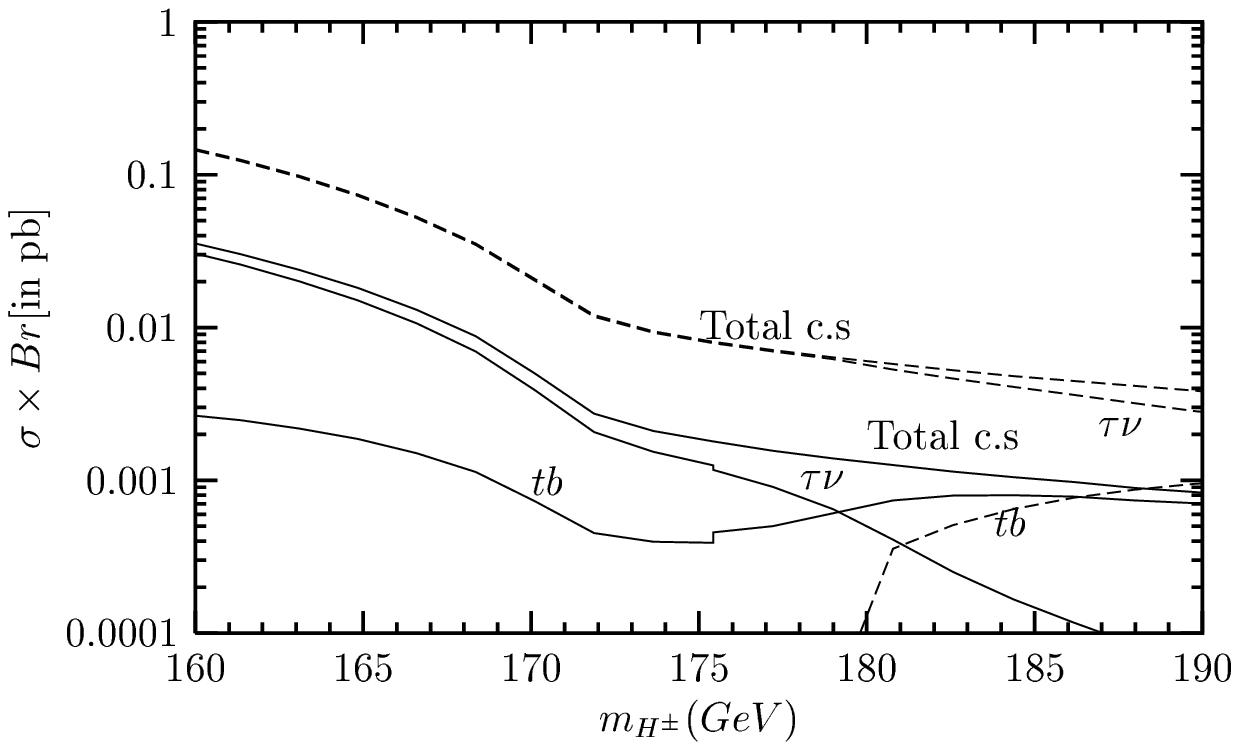}
\epsfxsize= 10.cm
\leavevmode
\hspace{-1.5in}
\epsffile{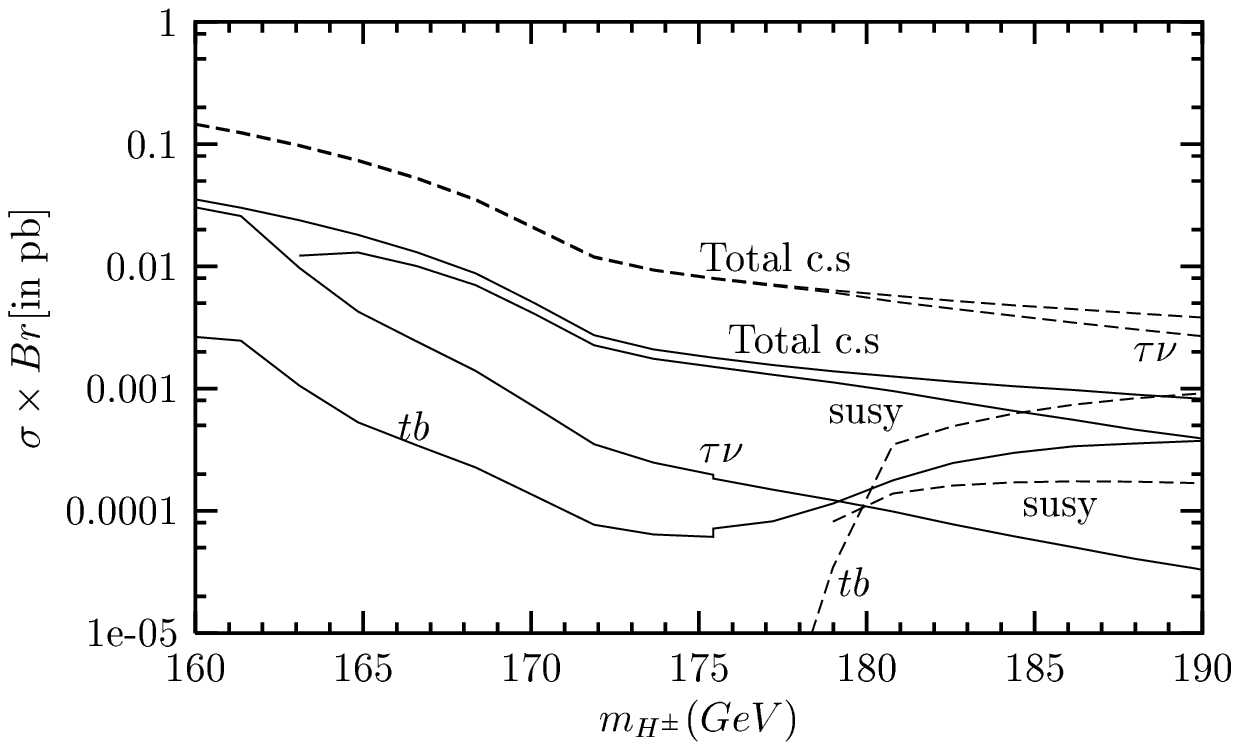}\\
\hspace{-0.5in}
\leavevmode
\vspace{-2.25in}
\caption{Cross section for $gg,q\bar q\to t\bar b H^-$ 
times the BRs in all relevant decay modes of charged Higgs
bosons, at $\sqrt s=2$ TeV,
as a function of $M_{H^\pm}$ for $\tan\beta
=3$ (solid) and 30 (dashed).
On the left-hand side, we assume
that decays into Supersymmetric particles are prohibited.
On the right-hand side, we include them, by adopting an
MSSM setup with $M_2=130$ GeV and $\mu=300$ GeV (see the text for
the sparticle masses).}
\label{fig_sigmaBRs}
}

\vfill\clearpage

\FIGURE[!t]{
\vspace*{-3.5cm}
\hspace*{-3.0cm}
\mbox{\psfig{file=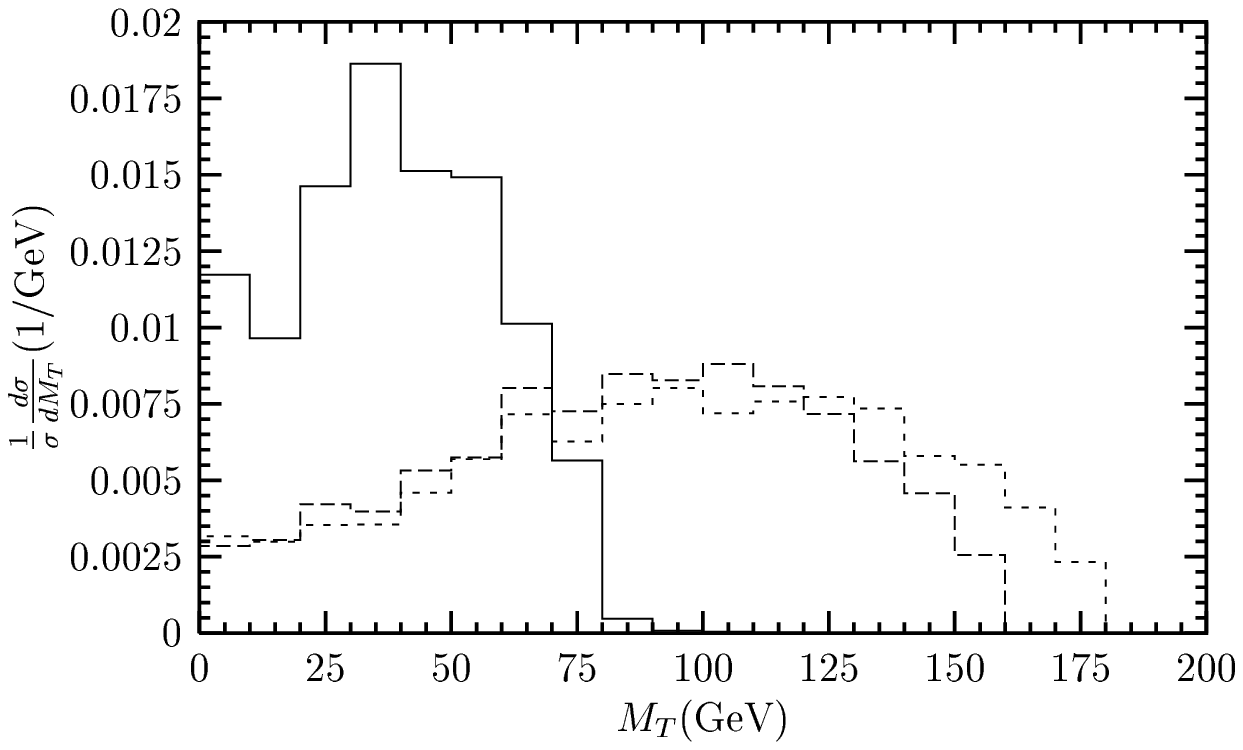,width=20cm}}
\vspace*{-13.0cm}
\caption{Transverse mass distribution, as defined in eq.~(\ref{mT}),
in $gg,q\bar q\to t\bar b H^-$, for $M_{H^\pm}$=160 and 180
GeV (long- and short-dashed, respectively), and 
$gg,q\bar q\to t\bar b W^-$ (solid), after the cuts 1.--4. 
described in the text. (Notice that the signal spectra are independent
of the choice of $\tan\beta$.)}
\label{fig_mT}
}

\vfill\clearpage

\TABLE[t]{
\begin{tabular}{|c|c|c|c||c|}
\hline
$M_{H^\pm}$ (GeV) $\downarrow$ \, / $\tan\beta$ $\rightarrow $  & 3  & 6 & 40 
& $t\bar b W^-$ \\
\hline
150 & 6   & 3   & 52&0.22  \\
160 & 2.8 & 1.5 & 22& 0.22 \\
170 & 0.4  & 0.25 & 3.5& 0.22\\
175 & 0.13 & 0.08 & 1.42& 0.22\\ 
180 & 0.067 & 0.061 & 1.09& 0.22\\
\hline
\end{tabular}
\vspace{0.5cm}
\caption{The cross section (in fb) for the signal 
$q\bar q,gg \rightarrow t \bar b  H^-(\rightarrow \tau^-
\bar\nu_{\tau})$ and the background 
$q\bar q,gg \rightarrow t \bar b  W^-(\rightarrow \tau^-
\bar\nu_{\tau})$, at $\sqrt s=2$ TeV,  for representative values of 
$M_{H^\pm}$ and $\tan\beta$, after all cuts described in the text.
 (Notice that the background rates are independent of
$M_{H^\pm}$, as the transverse mass constraint that
we adopted does not depend on the latter.)} 
\label{tab}
}

\vfill\clearpage
\FIGURE[!t]{
\vspace*{-3.5cm}
\hspace*{-3.0cm}
\mbox{\psfig{file=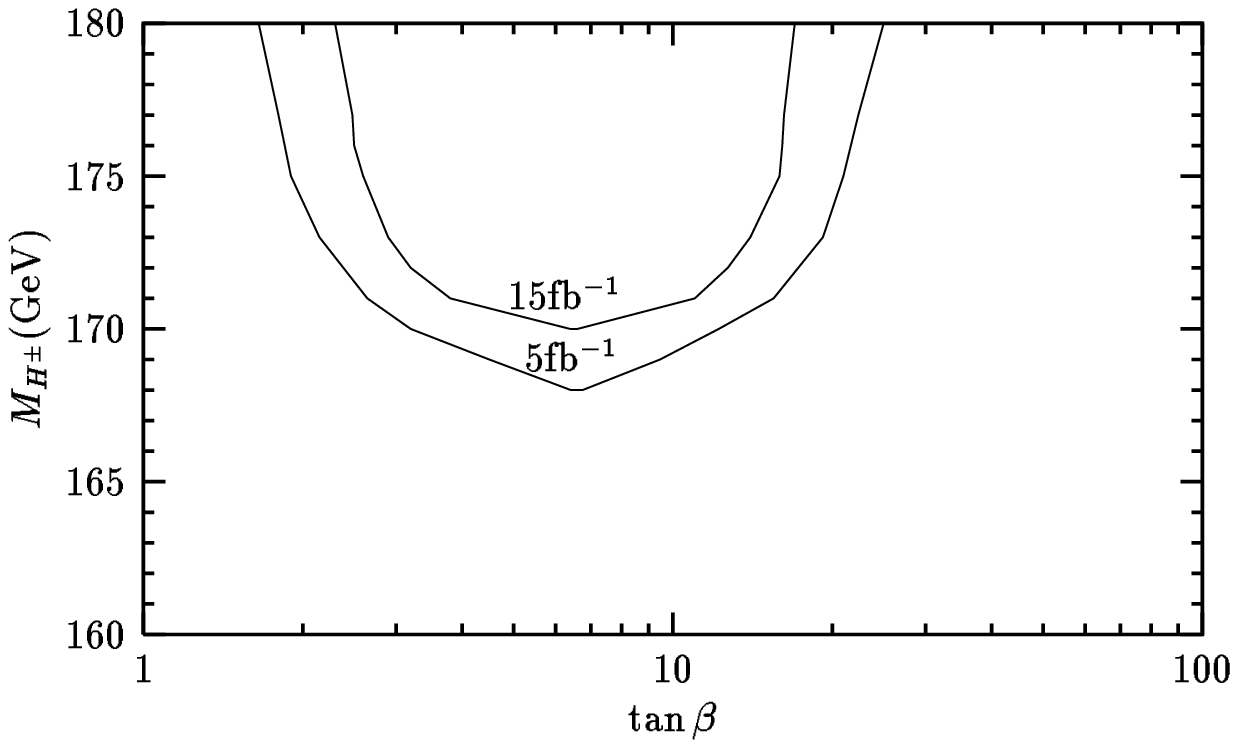,width=20cm}}
\vspace*{-13.0cm}
\caption{The exclusion regions (below the curves) at 95\% CL in the 
$M_{H^\pm}$--$\tan\beta$ plane that can be achieved at
Tevatron Run 2 for the two luminosity options of 5 and 15 fb$^{-1}$.}  
\label{fig_excl}
}


\begin{thebibliography}{MM}
\bibitem{dpcode} We have used a code provided by D.P. Roy to generate the 
distributions of the $\tau$-decay products, based on the formalism
discussed in \cite{dptau}.
\end{thebibliography}
\end{document}